# Speeding up of microstructure reconstruction: II. Application to patterns of poly-dispersed islands

W. Olchawa and R. Piasecki*,

*Institute of Physics, University of Opole, Oleska 48, 45-052 Opole, Poland*

HIGHLIGHTS

- A weighted doubly-hybrid method of statistical reconstruction of patterns is reported.
- The weighting factors account in part for an irregularity of islands shapes.
- A final pattern has the same value of interface and number of islands as the target.
- This competitive approach offers morphologically credible patterns.

ABSTRACT

We report a fast, efficient and credible statistical reconstruction of any two-phase patterns of islands of miscellaneous shapes and poly-dispersed in sizes. In the proposed multi-scale approach called a weighted doubly-hybrid, two different pairs of hybrid descriptors are used. As the first pair, we employ entropic quantifiers, while correlation functions are the second pair. Their competition allows considering a wider spectrum of morphological features. Instead of a standard random initial configuration, a synthetic one with the same number of islands as that of the target is created by a cellular automaton. This is the key point for speeding-up of microstructure reconstruction, making use of the simulated annealing technique. The program procedure allows requiring the same values for the reconstructed and target interface. The reconstruction terminates when three conditions related to the accuracy, interface and number of islands are fulfilled. We verify the approach on digitized images of a thin metallic film and a concrete sample cross-section. For a given accuracy, our method significantly reduces the number of accepted Monte Carlo steps when compared to the standard approach. At the same time, it provides credible shapes and similar areas of islands, keeping their number and the total interface of the target. To the best of our knowledge, this is the first attempt to obtain such an outcome. The cost-effective reconstructions suggest that the present technique could also be used for patterns of islands with strongly jagged border lines.

(Some figures may appear in colour only in the online journal)

Keywords: Multi-scale statistical reconstruction; Entropic descriptors; Correlation descriptors; Simulated annealing

* *E-mail address*: piaser@uni.opole.pl (R. Piasecki).

## 1. Introduction

In search of a practical use of novel or modified complex random multi-phase materials, prediction of their physical properties is the key point [1]. A description of the effective properties of heterogeneous materials involves microstructure/property connections and information related to phase volume fractions, particles shapes, their sizes and orientation, also interfaces, to mention but a few basic factors. A quick estimation of some typical macroscopic properties of random heterogeneous media, like thermal or electrical conductivity, can be obtained by means of one of the well-known standard effective medium approximations (EMA), the Maxwell-Garnet and the Bruggeman [2, 3], the differential effective medium theory (D-EMT) [4] or other extensions and generalizations [5]. A few simple modified lattice models show the impact of the model distribution of grains size and their shapes on the effective conductivity as well as the percolation threshold [6-8]. The effective transport properties of a paradigmatic structure composed of a dispersion of particles with real imperfect – multi-layered – structured interfaces can be effectively described, making use of the dual interface models, cf. [9] and citations therein. Nowadays, for the digitized media the interesting properties of a microstructure, e.g., the values of the phase interfaces can be directly extracted. This creates a lot of possibilities for the modelling of effective properties of the multi-phase media.

On the other hand, valuable microstructural information, though limited, can be obtained by correlation functions (CFs) which provide some details about the spatial distribution features. This kind of information is important for the stochastic reconstruction of a given microstructure by the simulated annealing (SA) technique within the Monte Carlo (MC) method [10]. A brief comparison of SA with genetic algorithm (GA) and maximum entropy (MaxEnt) optimization techniques is provided in Ref. [11]. It should be stressed that the concept of mixing of measures of different microstructural features is quite a popular technique utilized for reconstruction purposes. Recently, the two-point correlation and lineal path functions have been discussed from the point of view of their efficiency and accuracy of microstructure reconstruction [12]. Within the SA approach, particularly efficient is the hybrid pair of the standard two-point correlation function, $S_2$, providing information about the distribution of pair separations and the cluster one, $C_2$, sensitive to topological connectedness information [13]. These functions may be equivalently treated as *spatial correlation* descriptors. Some of the statistical descriptors or various structure metrics quantifying the topological connectivity can be exploited to optimize designed new materials, predict their tailored properties for specific applications and connect the material processing parameters with the resulting microstructure, or to consider various types of constraints for the reconstruction process [14-19].

The aim of this paper is different yet. We focus on optimizing the reconstruction process for wide-ranging microstructures composed of separated irregular and compact clusters. Why this topic deserves some attention? Obtaining, in an experiment, a long enough series of digitized micrographs of cross-sections for a given 3D sample is quite cumbersome and usually expensive. On the other hand: “In practice, one often has only a single digitized image to analyse (not an ensemble of them)” – see page 288 in [2]. But generation (for the sake of a further analysis) of a series of reconstructed reliable microstructures is quite cheap under the condition that the process is optimized. This is the main reason why we deal with the speeding up of microstructure reconstructions. Obviously, instead of the exact duplication of the target microstructure we focus on statistical approximate similarity between each of the final reconstructions and the parent microstructure. Thus, an unavoidable tolerance is connected with an assumed accuracy for a chosen MC technique.

Recently, for grey-scale patterns a hybrid reconstruction making use of SA algorithm suitably modified has been proposed [20]. It is based on a hybrid pair of the so-called *entropic* descriptors (EDs), the $S_\Delta$ (for the spatial inhomogeneity) and the $C_S$ (for the spatial statistical complexity), which were extensively described in Ref. [20]. Then, the approach was applied to speeding up of the reconstruction of a labyrinth microstructure [21]. This type of patterns belongs to a morphologically opposite class in comparison to patterns of poly-dispersed islands. The computational cost was quantified by the total number $N_t$ of required MC steps. We have found that the alternating random/biased reconstruction scenario reduces sufficiently the number $N_t$. However, the occurrence of some of the characteristic target’s morphological attributes in the synthetic initial pattern was the crucial factor in reduction of $N_t$. The starting pattern has been created by a cellular automaton that uses an activation/inhibition procedure described by Young [22]. As we have learnt recently, a similar idea of the beginning of a porous microstructure reconstruction with a synthetic initial configuration was applied in Ref. [23]. However, the configuration was prepared by a ballistic random sequential deposition of non-overlapping spherical particles. Recently, one more approach avoiding a random initial configuration within the graph-based SA reconstruction has been proposed and applied to a tomographic image of electrodes in Li-ion batteries [24].

The MC method applied in Ref. [21] makes use of only one hybrid pair, the $S_\Delta$ and the $C_S$. It was our first attempt to optimize SA reconstructing of the continuous two-phase media. It should be noted that the concept of using a combination of a pair of entropic and a pair of correlation descriptors, called here a doubly-hybrid approach, has already been mentioned earlier [20]. Now, we are in the position to state that such a cost-effective statistical reconstruction of the discontinuous media is viable. The program of this kind is more capable of capturing distinct morphological features of different microstructures. Two examples of such two-phase microstructures are given in Section 5. One of them is Co/C thin film evolving along temperature [25] while a concrete sample cross-section considered in Ref. [13] is the second one. Now, instead of the total number $N_t$ we prefer to use the number $N_a$ of the accepted MC steps. We have found, it is usually about thirty percent of the $N_t$.

In this work, we report the fresh and competitive *weighted doubly-hybrid* (WDH) method of the statistical reconstruction that uses simultaneously a pair of entropic descriptors $\{S_\Delta, C_S\}$ and a pair of correlation functions $\{S_2, C_2\}$. The details of preparing synthetic initial configurations by a chosen cellular automaton are given in Section 2. Our main program is described in Section 3. It includes a considerably modified procedure of switching between weak/strong bias modes during the reconstruction process. As a result, the current value of the two-phase interface can be kept under control. The utility of monitoring of the interface has been shown using the examples of islands of miscellaneous shapes and poly-dispersed in sizes, mentioned above [13,25]. It is important to emphasize that in the present approach, a competition appears between the two pairs of the considered statistical descriptors. In this way, dissimilar/complementary morphological features of a target can also be taken into account, at least in some part. However, the attainability of an assumed accuracy by the WDH approach can be a challenging task. Thus, it is highly desirable to improve and speed up this method. In Section 4, we exemplify the effectiveness of our method. The results obtained suggest the meaningful reduction of the needed final number of the accepted MC steps when compared to the standard reconstruction starting with a random initial pattern.

## 2. Creation of clusters

We need to generate a set of $N$ synthetic islands (compact clusters) built of black pixels, treated as unit cells $1 \times 1$, and with specified areas $A_1, A_2, \ldots, A_N$. The clusters should be placed randomly at $L \times L$ matrix of white pixels. In order to perform such a procedure, two different approaches within cellular automata frame are developed. The first automaton (CA1) uses certain rules connected with the Gaussian distribution, while the second one (CA2) employs a simple probabilistic approach. However, in both cases the procedure of drawing of the "centres" $P_1, P_2 \ldots$ and $P_N$ for the $N$ clusters is a common part. It uses the same two arbitrary chosen parameters $d$ and $a$, within intervals $0 \le d < d_{\max}$ (typically $d_{\max} \approx 4$) and $0 < a < 1$, see (2.2) and (2.3), respectively. The auxiliary radius $r_i$ of a circle attributed to each of the $N$ centres is given by

$$r_i = \sqrt{A_i / \pi}\,. \tag{2.1}$$

In addition, the following additional constraints should be satisfied:

$$\rho(P_i, P_j) > r_i + r_j + d\,, \tag{2.2}$$

where $\rho$ is a distance between any two points $P_i, P_j$ and

$$\rho_0 / r_i > a \tag{2.3}$$

for each point $P_i$ at a distance $\rho_0$ from an edge of the generated pattern. The further proceeding of generation of clusters is different for each of the considered models.

First, we present some details for the more complicated model CA1. This model makes use of additional parameters $0 < b_k < 1$, $k \in \{1, 2, 3\}$, which are specific to the model. To make the way it works clearer, we introduce two coordinate systems: (*i*) the global one ($X$, $Y$) with axes parallel to the edges of the considered pattern and passing through the centres of pixels; (*ii*) the local one $(x, y)$ having a shifted origin and rotated axes according to the procedure given below. In a local coordinate system different for each $P_i$, the positions of points are generated using two-dimensional Gauss distribution

$$f(x, y) = \frac{1}{2\pi\sigma_1\sigma_2} \exp\left[ -\frac{1}{2}\left( \frac{x^2}{\sigma_1^2} + \frac{y^2}{\sigma_2^2} \right) \right]. \tag{2.4}$$

Then, making the transformation from a local to global description, we obtain the coordinates

$$\begin{aligned} X &= X_i + x\cos\beta_i + y\sin\beta_i \\ Y &= Y_i - x\sin\beta_i + y\cos\beta_i \end{aligned}\,. \tag{2.5}$$

For each cluster the two random variables $\sigma_1$ and $\beta_i$ are generated according to the formulas

$$\sigma_1 = b_2\{1 + b_1(rand - 0.5)\}\, r_i\,, \tag{2.6}$$

$$\beta_i = \pi\ rand\,, \tag{2.7}$$

where *"rand"* means random variable of uniformly distributed on interval [0, 1). In turn, the parameters $b_1$ and $b_2$ (typically $b_2 \approx 0.4$) enable us to create elliptic shapes of clusters. Then the variable $\sigma_2$ is given by

$$\sigma_2 = b_2 A_i / \pi\sigma_1 \, . \tag{2.8}$$

After rounding ($X$, $Y$) to the nearest integer values, the position of a pixel is determined. If the pixel is colour white then it is changed to the black one. The counted number of black pixels becomes the current value of the $i$th cluster area. However, when the ratio of current area $A_i'$ to the target one $A_i$ of the growing cluster exceeds a fixed value of parameter $b_3$,

$$A_i' / A_i > b_3 \, , \tag{2.9}$$

the black pixel is attached conditionally. Namely, subject to the condition, it has at least one nearest neighbour (n.n.) that belongs to the growing cluster. In this way we avoid the appearance of undesirable cluster spreading. The procedure is finished when the number of black pixels equals a given cluster area and repeated, when successive clusters are generated.

Now, we briefly describe the second model CA2. During the growing of any of the black clusters, a random selection of white pixels is specified to the cluster's surrounding only. Thus, a drawn white pixel can belong to only one of the following groups: (a) *active surroundings* composed of white pixels being n.n. of a single cluster only, (b) *passive surroundings* consisting of those white pixels, which are n.n. at least of two clusters. Let us describe the initial stage of the growing process of one pixel cluster. The von Neumann neighbourhood is taken into account in the next step. To attach a second black pixel, a position among the active surroundings should be randomly drawn. We use the arbitrarily chosen set of fixed probabilities $p_l$ of any black pixel attachment, $0 \le p_1 \le p_2 \le p_3 \le p_4 \le 1$. The index $l$ indicates the number of black n.n. of a pixel of the active surroundings. After the acceptance, in this case with probability $p_1$, we have a two-pixel cluster and six positions of the new active surroundings. The growing process terminates when the needed area of the cluster is reached. In this way, by choosing appropriate parameters $p_l$, quite regular shapes of almost compact clusters can be obtained.

It should also be noticed that in a single step of the growing process, the status of a white pixel belonging simultaneously to the surroundings of two clusters cannot change. Hence, in contrast to the CA1 approach, the merging of clusters cannot be observed in the CA2 model. On the other hand, within the former approach one can force the creation of elliptically shaped and flattened clusters, while in the other model such an option is not considered. It allows choosing a more suitable model, CA1 or CA2, depending on specific features of poly-dispersed islands. Finally, one can obtain a preliminary series of trial configurations with the target's basic attributes, such as the number and areas of clusters. Then, one of them for which the statistical "distance" from the target curves – given in the next section by the objective function (3.3) – is the smallest one, becomes the synthetic starting pattern for microstructure reconstruction.

## 3. Weighted doubly-hybrid reconstruction

The simulated annealing (SA) technique is frequently applied to reconstruct a microstructure. When the entropic descriptors (EDs) only are included in the objective (cost) function, the speeding-up of a labyrinth microstructure reconstruction can be achieved in a simple way [21]. Here, we propose a significant modification of this kind of reconstruction in order to apply it to patterns of islands of miscellaneous shapes and poly-dispersed in sizes. Now, besides a hybrid pair of two different EDs, we use an additional pair of quantifiers, namely a hybrid pair of distinct correlation functions (CFs). The basic definitions of the EDs utilized in Refs. [20,21] are given in Appendix. In particular, the $S_{\Delta}$ quantifies *spatial* inhomogeneity, while the $C_S$ relates to *spatial* statistical complexity at different length scales. The second exploited hybrid pair comprises two-point correlation function $S_2$ and two-point cluster function $C_2$, fully described, e.g., in Ref. [2]. The effectiveness of the latter pair has recently been clearly demonstrated for textures taken from materials science, cosmology, and granular media [13]. The authors underline that the addition of $C_2$ is of the key importance to improvement of the reconstruction process compared to the usage of $S_2$ alone. To calculate $C_2$ we have applied a method of labelling of clusters implemented in the algorithm of Newman and Ziff [26].

It is worth mentioning that the two competing hybrid pairs are of different origin and they appear with weighting factors in the final formula for the objective function (3.3) given below. Thus, the full name of the present approach is the weighted doubly-hybrid (WDH) method. We expect that the competition of the pairs will allow considering a wider spectrum of morphological features. For an adapted target's pattern of size $L \times L$ in pixels the modified objective function can be described as average "energy" per a descriptor. Now, the objective multi-scale function is the weighted sum of squared and normalized differences between the values of binary EDs related to the current configuration and the target pattern for the black phase, and similarly, between the values of the CFs. For the purpose of making a comparison, the energy is additionally averaged over the number $n$ of considered length scales. To obtain a well-balanced influence of ED and CF part in (3.3), the equal number of considered scales is applied. Let us note that the present form of energy differs from the previously considered one [20,21], where two pairs of binary entropic descriptor and its grey-scale counterpart are included.

The differences are normalized with respect to the maximal values of target EDs and CFs marked with the superscript '0'. Correspondingly, the normalized EDs differences can be written as

$$\tilde{S}_{\Delta}(k) - \tilde{S}^{0}_{\Delta}(k) \equiv [S_{\Delta}(k) - S^{0}_{\Delta}(k)]/\max S^{0}_{\Delta}(k) \ , \tag{3.1}$$

$$\tilde{C}_{S}(k) - \tilde{C}^{0}_{S}(k) \equiv [C_{S}(k) - C^{0}_{S}(k)]/\max C^{0}_{S}(k) \ . \tag{3.2}$$

In a similar way, the related differences can be written for the CFs. The formula describing the final form of energy $E$ reads

$$E = \frac{1}{4n}\left\{ \alpha \sum_{k\,odd}^{L} \left[ \left(\tilde{S}_{\Delta}(k) - \tilde{S}^{0}_{\Delta}(k)\right)^2 + \left(\tilde{C}_{S}(k) - \tilde{C}^{0}_{S}(k)\right)^2 \right] + \right. \\ \left. (1-\alpha) \sum_{r=0}^{L/2-1} \left[ \left(\tilde{S}_{2}(r) - \tilde{S}^{0}_{2}(r)\right)^2 + \left(\tilde{C}_{2}(r) - \tilde{C}^{0}_{2}(r)\right)^2 \right] \right\} \tag{3.3}$$

Here, the parameter $0<\alpha<1$ and the two coefficients, $\alpha$ and $1-\alpha$, are treated as the weighting factors. Note that for the EDs and CFs the identical number $n = L/2$ of length scales appears.

To minimize the given energy we use a general scheme of MC method. For a current pattern, two randomly selected pixels of different phases are interchanged under certain conditions (a weak/strong bias mode), which are explained later on. The new trial configuration (the system's state) is then accepted with probability $p(\Delta E)$, according to the Metropolis-MC acceptance rule [10]

$$p(\Delta E) = \begin{cases} 1 & \Delta E \le 0, \\ \exp(-\Delta E/T) & \Delta E > 0, \end{cases} \tag{3.4}$$

where $\Delta E = E_{\text{new}} - E_{\text{old}}$ is the change in energy between the two successive states. Upon acceptance, the trial pattern becomes a current one, and the evolving procedure is repeated for a given current loop's length. A fictitious temperature $T$ changes, following the popular cooling schedule $T(l)/T(0) = \gamma$ with the chosen parameter $\gamma = 0.85$ and initial temperature $T(0) = 10^{-7}$. The $l$ means the $l$th temperature loop associated with the annealing process.

A few words are needed on the modifications of the exchange procedure used previously [21]. In order to optimize the weak bias mode (WBM) drawing procedure, we should avoid selecting specific pairs of different colour pixels. To be precise, neither the white pixel of passive surroundings (see Section 2) nor the black one of a cluster's interior is allowed. Such a restriction leads commonly to an increase in the interface value. In our program, a pair of white and black pixels is accepted by WBM drawing procedure when the simplifying conditions are simultaneously satisfied: (a) a white pixel is a member of active surroundings; (b) a black pixel belongs to the cluster's surface called temporarily a border.

Denoting the numbers of black n.n. and n.n.n. for a white centre as $w_{nn}$ and $w_{nnn}$, and similarly, for a black centre as $b_{nn}$ and $b_{nnn}$, we can describe, in a simple way, the strong bias mode (SBM). This mode can be applied even at the first step of the drawing procedure during the reconstruction. The two pixels of different phases selected randomly within the SBM must fulfil both the WBM conditions and the additional constraints, which are specific to the SBM:

$$(b_{nn} + b_{nnn} < w_{nn} + w_{nnn}) \quad \text{and} \quad (b_{nn} \le w_{nn}) \tag{3.5}$$

or

$$(b_{nn} + b_{nnn} = w_{nn} + w_{nnn}) \quad \text{and} \quad (b_{nn} < w_{nn})\,. \tag{3.6}$$

Contrary to the WBM, the conditional exchange of a white and black pixel within the SBM favours a decrease in the interface value. It is worth mentioning that instead of the two conditions (3.5-6), the less demanding inequality can be used, $b_{nn} + b_{nnn} \le w_{nn} + w_{nnn}$, leading to roughly similar results. However, the latter condition usually increases the total number $N_a$ of the accepted MC steps, when the same accuracy of a reconstruction is assumed.

To control the value of current interface $I_{current}$ by an algorithm, two monitoring mechanisms have been implemented. The first of them is the tracing routine. For this purpose, the value of $I_{current}$ is kept within a properly chosen interval, $[I_{target} - \eta_{max}, I_{target} + \eta_{max}]$. This interval is successively shrunk, according to an established scenario. Namely, the interval's margins result from the varying maximal deviation, $\eta_{max} = \max(\eta_1, \eta_2)$. The first auxiliary deviation $\eta_1$ is described by a simple relation

$$\eta_1 = \begin{cases} \eta_0 & \text{for} \quad l \le l_S \\ \eta_0 - (l - l_S)\mu & \text{for} \quad l > l_S \end{cases} \tag{3.7}$$

where $\eta_0$ is a given initial value of $\eta_1$, $l_s$ is the first temperature loop with the non-zero constant shrinking $\mu$-parameter, for example $\mu = 2$. The second auxiliary deviation $\eta_2$ is determined by

$$\eta_2 = \left| I_{\text{current}} - I_{\text{target}} \right| . \tag{3.8}$$

If a MC step changes the value of $I_{\text{current}}$ in such a way that $I_{\text{current}} \notin [I_{\text{target}} - \eta_{\max}, I_{\text{target}} + \eta_{\max}]$ then this step is ignored.

The second mechanism controlling the value of $I_{\text{current}}$ consists in switching of the procedure of the pixels exchange between the WBM/SBM described above. We remind that usually the SBM supports a decrease in the $I_{\text{current}}$ value, while the WBM favours its growth. Thus we apply the following rule of the switching: if the $I_{\text{current}}$ value exceeds the value of $I_{\text{target}}$, the SBM comes into play while in the opposite case the WBM becomes active.

The whole simulation runs with the increasing length $N(l)$ of successive temperature $l$-loops. The length $N(l)$ is an integer part of a simple function $f(l)$ given by

$$f(l) = A + (l-1)B + C^{(l-1)/l^*} - 1 \ . \tag{3.9}$$

For the $l > l^* = 23$ the length increase becomes more and more non-linear. The other coefficients $A = 175$, $B = 35$ and $C = 400$ correspond to the length of the initial loop, the linear part of the increase of loop's length and the base of power function, respectively. In most cases of the performed simulations, the overall number of temperature loops was smaller than 26.

To finish the evolving procedure, the following conditions are considered. When the energy $E$ becomes smaller than the assumed tolerance $\delta$-value, in our case $\delta = 10^{-5}$, then we touch the first condition F1. Further, the WDH method which enables us to trace the current value of a two-phase interface allows requiring the same values for the reconstructed and target interface. Thus, the next condition F2, $I_{\text{final}} = I_{\text{target}}$, come into play. As the last involved condition F3, we choose the conservation of the number of islands, i.e. $N_{\text{isl, final}} = N_{\text{isl, target}}$. If the three conditions are satisfied then the WDH reconstruction terminates.

## 4. Examples of Monte Carlo simulations

Our aim is to speed up the microstructure reconstruction of patterns of poly-dispersed islands. However, besides the standard finishing condition F1 related to the accuracy of the reconstruction, the final patterns should satisfy the additional conditions, F2 and F3. They are connected with the conservation of the values of interface and the number of islands. This is a reason why, for the sake of an efficient reconstruction, instead of a random initial configuration, we use its synthetic counterpart. This pattern is suitably generated by one of the cellular automata mentioned earlier. In this way, the characteristic attributes of a given target are taken into account at the initial stage. This is one of the key factors in reducing the number $N_a$ of the accepted MC steps for fixed reconstruction accuracy.

On the other hand, for the first time we have applied a combination of two hybrid approaches. The first approach is based on two entropic descriptors, while the other one uses a pair of two-point correlation functions. Such an approach permits to optimize the MC reconstruction by means of the weighting $\alpha$-parameter in the objective function given by (3.3). We verify our approach on exemplary two-phase microstructures: (1) thin film with the poly-dispersed metallic islands adapted from [25], and (2) concrete cross-section with the irregularly shaped stone phase adapted from [13].

Below, the same main scenario for the SA simulations has been used. For the chosen values of the weighting parameter, $\alpha = 0.1, 0.2\ldots\ 0.9$, a series of 20 WDH reconstructions were performed. Each of them utilized a different random seed but the same specific initial configuration, the pattern Synth1 (for Example 1) and the pattern Synth2 (for Example 2). For each of these examples, a series of 1000 trial energies has been provided by an appropriate cellular automaton on the basis of two different lists of numbers of the target's islands and their sizes. In both cases, to compute trial energy, the middle value $\alpha = 0.5$ is used. The two selected configurations had the smallest value of the corresponding energy described by Formula (3.3).

### 4.1 Example 1

First, we examine a system of quite regular (in shapes) metallic islands for thin cobalt films of 2 nm thickness. They were fabricated by evaporation of pure cobalt under high-vacuum conditions ($10^{-6}$ Pa) onto the platinum microscope grids covered with amorphous carbon substrate. The thin film carbon substrates were prepared by vacuum deposition from a carbon arc. The investigated Co/C film was heated in flowing purified $H_2$ at 873 K for 4h. In digitized transmission electron micrograph (TEM) of size $1080 \times 1080$ in pixels the length of 1 cm corresponds to 1000 Å. The micrograph's resolution was decreased twice. Then, five sub-domains of size $180 \times 180$ of equal cobalt surface coverage $\varphi = 0.17$ (cobalt concentration closest to that of the micrograph) were selected. The next two conditions were related to finding among them such patterns that show the first maximum of $S_\Delta$ and $C_S$ as near as possible to the corresponding length scales for the reduced micrograph. In this way, the best one serves as the target image T1 with the two-phase interface $I_{target}(T1) = 1660$ and the number of islands $N_{isl,\ target}(T1) = 32$; see Fig. 1a.

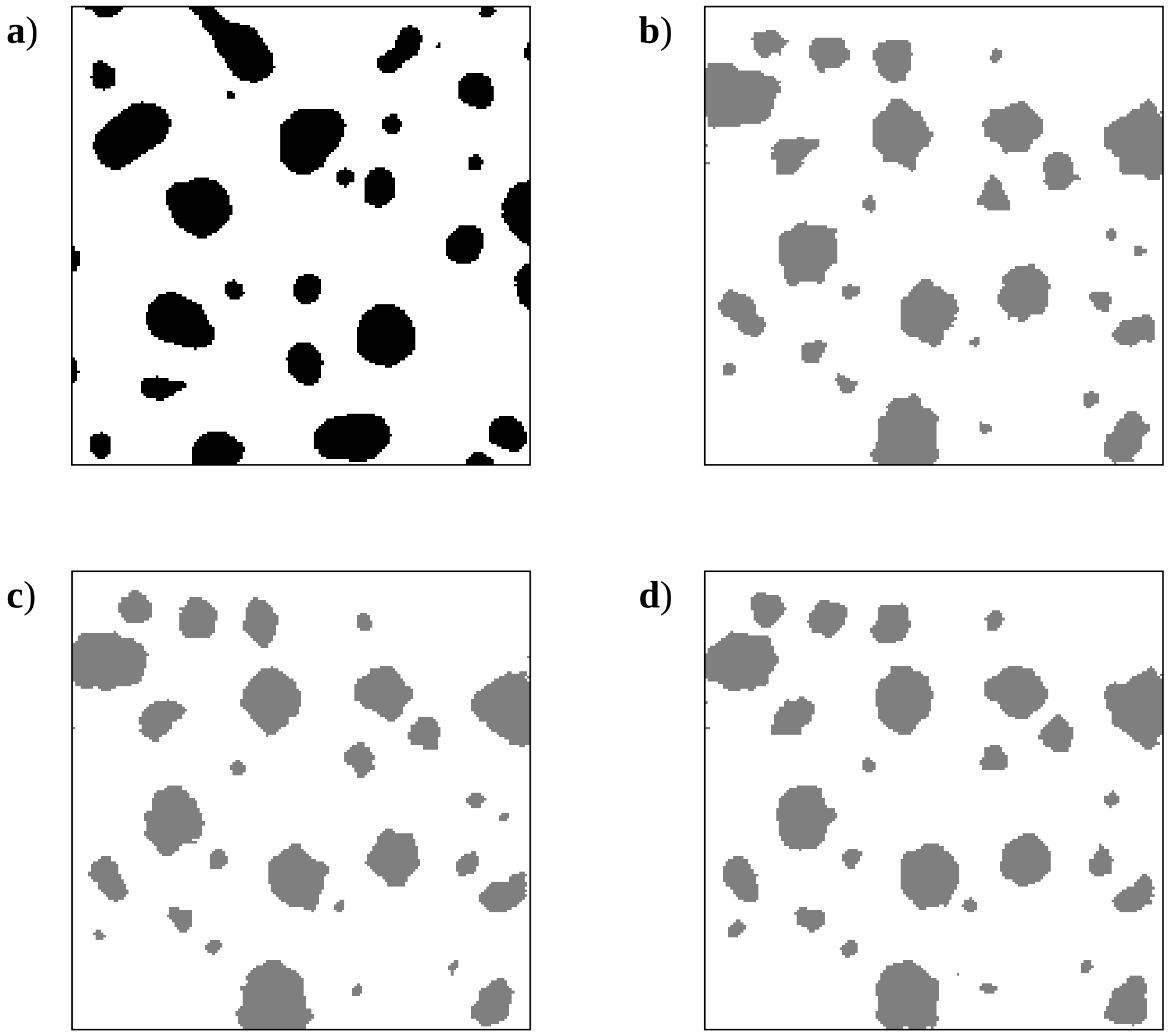

**Figure 1.** A comparison of the representative $180 \times 180$ digitized sub-domain of the poly-dispersed metallic islands adapted from Ref. [25] with its typical WDH reconstructions for a given random seed and chosen $\alpha$-parameters. a) Target image T1; b) The reconstruction with $\alpha = 0.1$; c) $\alpha = 0.5$; d) $\alpha = 0.9$.

At this stage, we are ready to test the WDH reconstruction on this representative arrangement of quite smooth islands. We start with the synthetic initial pattern Synth1 highlighted in the inset in Fig. 2a. Because of the low concentration of the black phase, the synthetic pattern is generated by the cellular automaton CA1 described in Section 2, which is more suitable for this case. The starting synthetic pattern Synth1 has the same number of islands as the target T1, while the value of the initial interface is slightly higher, i.e., $I_{initial}(\text{Synth1}) = 1713 > 1660$. As far as the characteristic positions of the first maximum are concerned, we are in a comfortable situation, since $k_{max}(S_{\Delta}; \text{Synth1}) = k_{max}(S_{\Delta}; \text{T1}) = 26$ and $k_{max}(C_S; \text{Synth1}) = 32$ slightly differs from $k_{max}(C_S; \text{T1}) = 31$. For one of the used random seeds, the typical reconstructions with the bottom, middle and top value of the $\alpha$-parameter are shown in Figs. 1b-d. One can see that the higher the $\alpha$ is, the smoother the islands shapes appear in the reconstructed patterns.

**2a**)

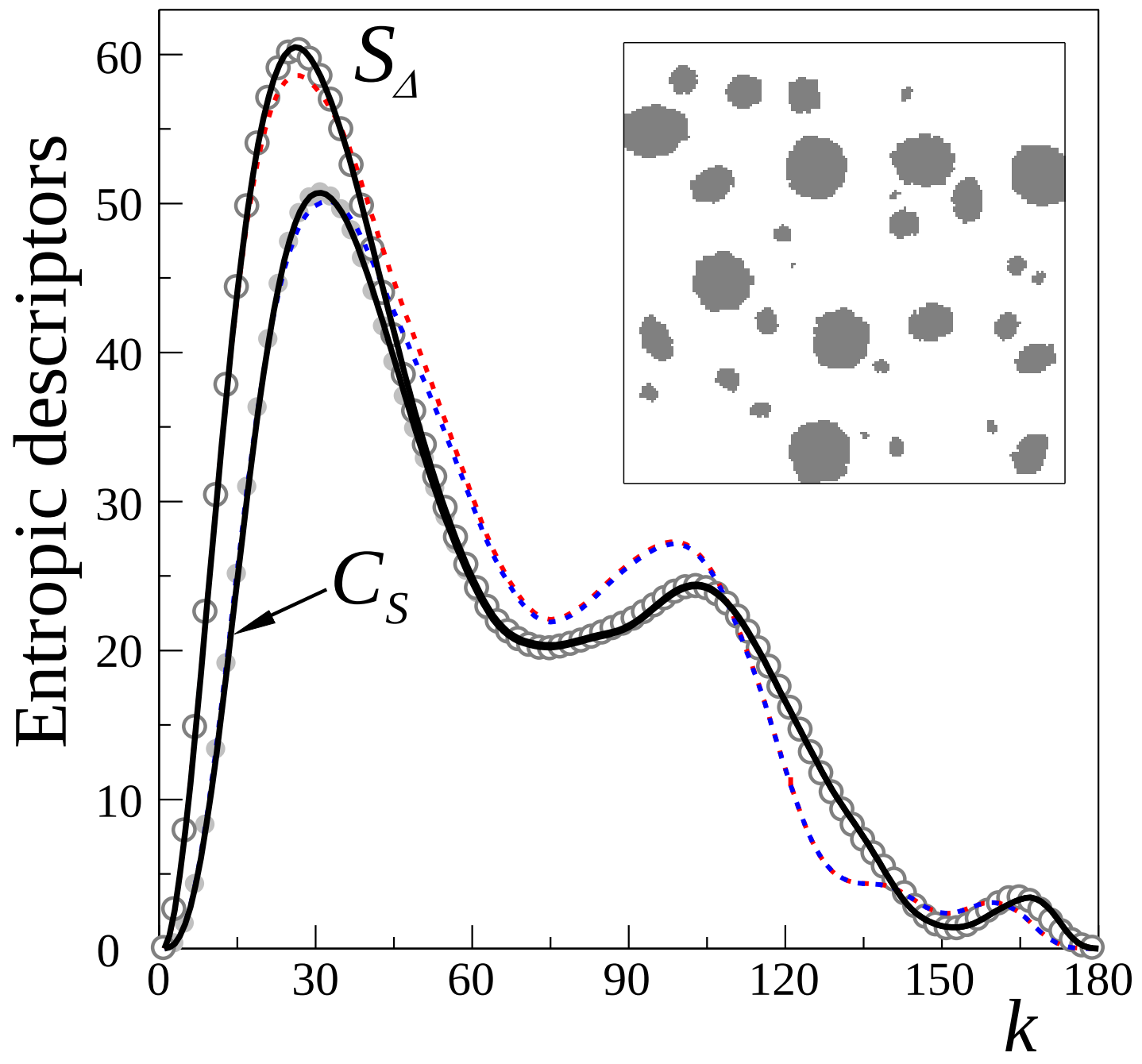

**2b**)

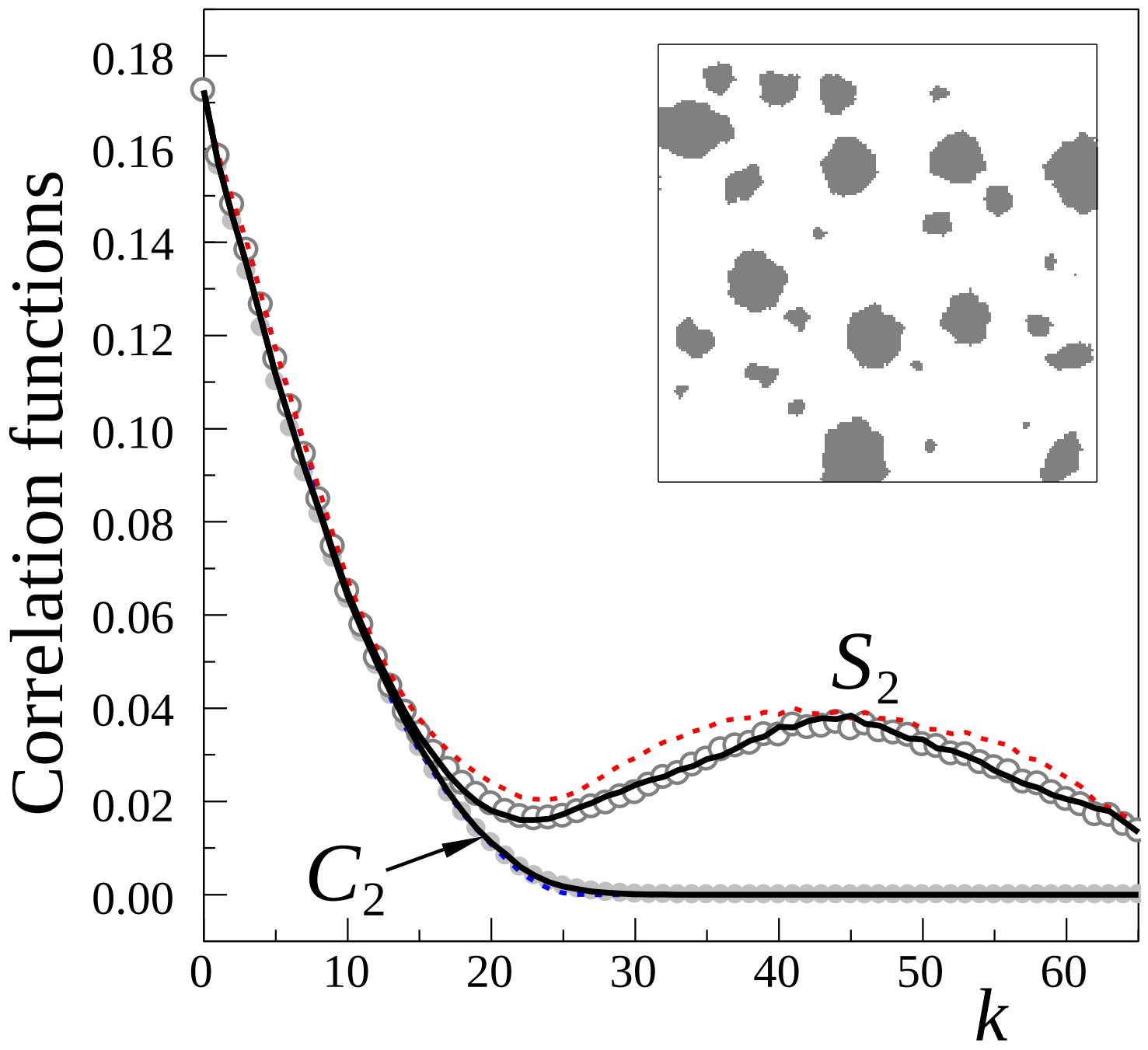

**Figure 2.** (Colour online) The quality of weighted doubly-hybrid (WDH) method shown using the example of the reconstruction R1*($\alpha = 0.5$; T1) selected as having the closest $N_a(\alpha = 0.5)$ value of the accepted MC steps to their average $<N_a(\alpha = 0.5)>$ over the series of 20 runs. a) For the T1-target, thick solid black lines, and for its R1*-reconstruction, open and filled grey circles, relate to the pair of entropic descriptors $\{S_\Delta; C_S\}$. For completeness, the dashed lines, red and blue online, describe the computed starting values for initial Synth1-pattern that is shown in the inset; b) The same correspondence for the pair of correlation functions $\{S_2; C_2\}$ but now the R1*-reconstruction is depicted in the inset.

In turn, Figs. 2a-b show the quality of one of the 20 reconstructions for the middle value of $\alpha = 0.5$. The related reconstruction R1*($\alpha = 0.5$; T1), depicted in the inset of Fig. 2b, has the closest $N_a(\alpha = 0.5)$ value to the average $\langle N_a(\alpha = 0.5)\rangle$ of the accepted MC steps. In Fig. 2a, the thick solid black lines (open and filled grey circles) correspond to entropic descriptors of spatial inhomogeneity and statistical complexity computed for the target T1 (its reconstruction R1*), respectively. For completeness, the EDs computed for the initial Synth1 given in the inset, are described by the dashed lines (red and blue online). Correspondingly, in Fig. 2b, the results for the correlation functions CFs are presented. Let us note that the curves for the initial synthetic images are qualitatively similar and relatively close to the target ones. This is connected with the inclusion of some of the target's characteristic attributes to the creation rules of a synthetic pattern by the chosen cellular automaton. At all scales $k$, a significant fit of the considered hybrid descriptors can be observed. The common small tolerance $\delta$-value forces an appearance of the equal quality fitting for the other random seeds, too.

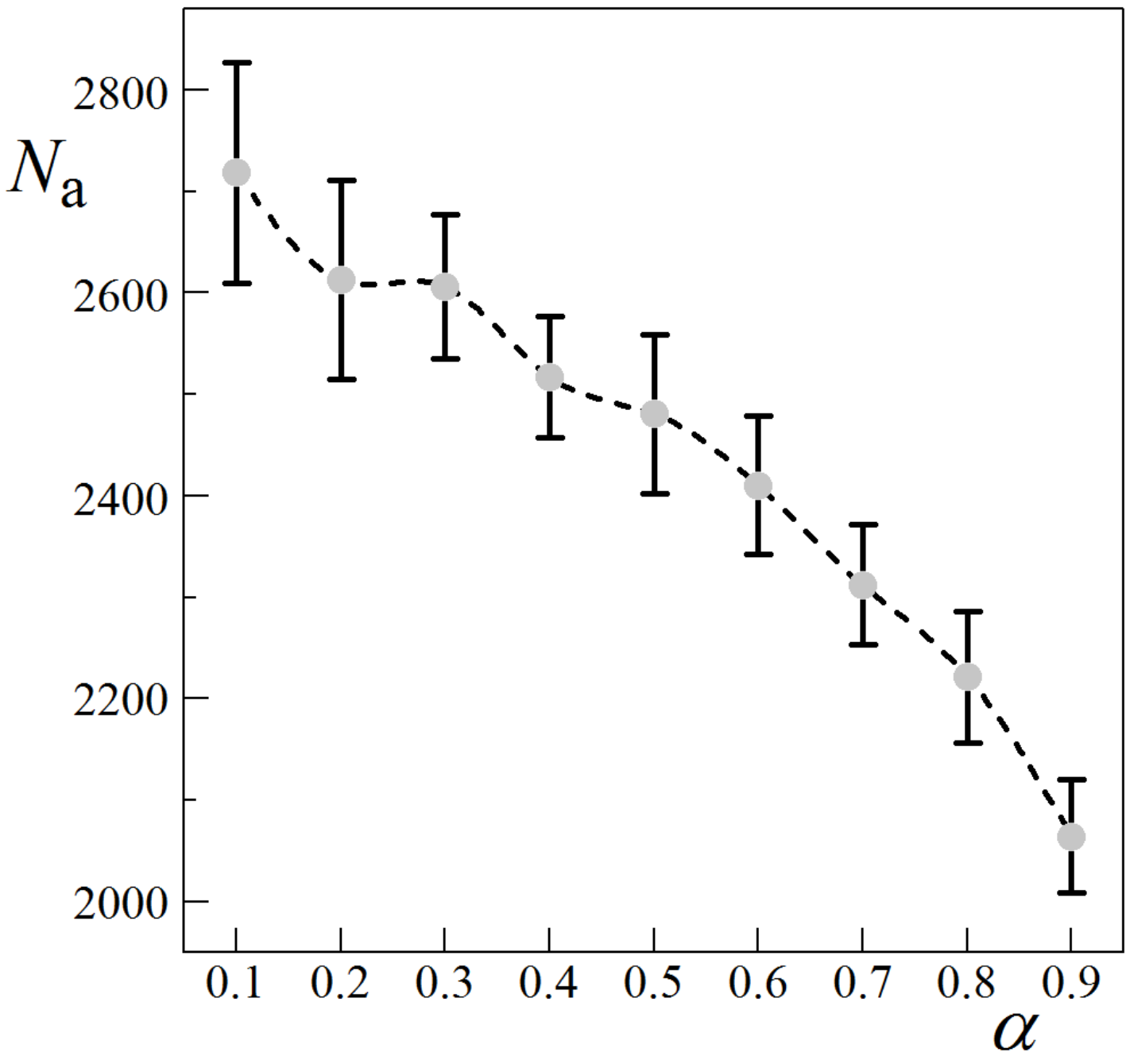

**Figure 3.** The statistics of $N_a(\alpha)$ accepted MC steps for the WDH reconstructions of the T1-target (given in Fig. 1a). The Synth1-pattern depicted in the inset of Fig. 2a is the initial configuration. For each of the fixed nine values of $\alpha$-parameters a series of 20 runs with given different random seeds was completed. The error bars represent the corresponding standard deviations. The dashed line connecting the averages $\langle N_a(\alpha)\rangle$ marked by the filled circles is a guide for eyes only. The shape of the curve is not a universal one. Nevertheless, it shows the generally decreasing trend in the averages, which belong roughly to the range from 2000 to 3000.

For each of the nine $\alpha$-parameters the series of 20 runs has been completed. The statistics of the results are illustrated in Fig. 3. The dashed line connecting the averages $<N_a(\alpha)>$ is a guide for eyes only. Although the shape of the curve is not a universal one, it shows a characteristic decreasing trend in the $<N_a(\alpha)>$ belonging roughly to the range from 2000 to 3000. It should also be mentioned that a standard MC reconstruction starting with a random initial configuration but using only the first terminating condition F1 was stopped after about $N_a = 10^5$ steps and still the obtained accuracy of the reconstruction was not satisfactory. Thus, our approach allows a meaningful reduction of the accepted MC steps at least for the case of islands which are quite regular in their shapes.

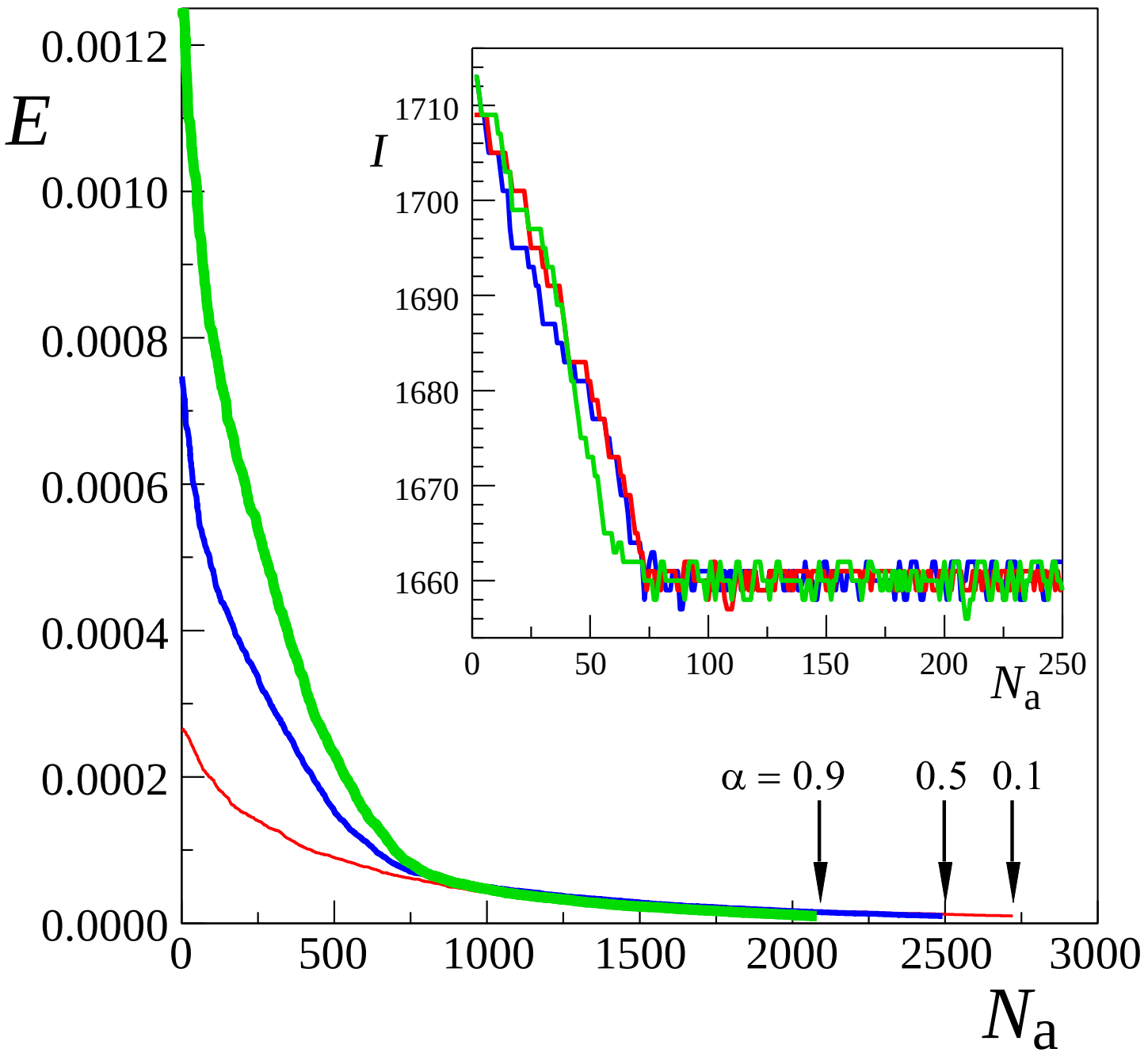

**Figure 4.** (Colour online) The convergence of energy $E$ as a function of the accepted Monte Carlo attempts $N_a$ for three representative cases: $\alpha = 0.9$ (thick green), 0.5 (thin blue) and 0.1 (the thinner red line) from the top to the bottom, which are the closest to the averages $<N_a(\alpha)>$. The arrows indicate the corresponding final values of $N_a(\alpha)$, which are obtained when the assumed tolerance $\delta = 10^{-5}$ is achieved and the remaining two conditions (see the text) are fulfilled. To illustrate the limitative mechanism for the current interface value, the dependence of the $I_{current}(\alpha)$ on the $N_a(\alpha)$ is shown in the inset. Actually, for $N_a > 70$ one can observe only interface fluctuations around the $I_{target} = 1660$.

In Fig. 4, the three $\alpha$-cases (given below) and the closest to the averages $<N_a(\alpha)>$ clearly demonstrate the fast convergence of the WDH reconstruction. The curves from the top to the bottom (the thick green line for $\alpha = 0.9$, medium blue for $\alpha = 0.5$ and thin red for $\alpha = 0.1$ in colour on-line) correspond to the absolute values of the energy given by (3.3). The arrows indicate the corresponding final values of $N_a(\alpha)$. Despite some understandable initial differences in energies $E$, they become negligible when $N_a(\alpha) > 1000$ for each of the $\alpha$-parameters. In order to illustrate in action the "snake-like procedure" that forces the bounding of interface value to the target one, the dependence of $I_{current}(\alpha)$ on the number $N_a(\alpha)$ is shown in the inset of Fig. 4. Actually, the dependence on $\alpha$-parameter quickly becomes not essential since for $N_a > 70$ the values of $I_{current}$ stay inside the narrow range around the $I_{target}(T1) = 1660$.

## 4.2 Example 2

For comparison, we examine a concrete microstructure adapted from Ref. [13]. Fig. 5a shows a binarized concrete sample cross-section being our target image T2 of size $170 \times 170$ in pixels. The white phase corresponds to the cement paste. The black phase of concentration $\varphi = 0.51$ represents the stones. For the adapted target image T2 one can find that the two-phase interface is $I_{target}(T2) = 3252$ and the number of black clusters (islands) equals $N_{isl,\,target}(T2) = 42$; see Fig. 5a. It was noticed that densely dispersed islands, irregularly shaped and of various sizes, make the statistical reconstruction of the T2 a nontrivial task [13].

Our WDH-reconstruction of such a highly non-uniform arrangement of irregular islands, which are relatively big in comparison to the size of the whole pattern, can be compared successfully with the original one presented in (Fig. 2d of Ref. [13]). We begin with the synthetic initial pattern Synth2 highlighted in the inset of Fig. 6a. This time, the cellular automaton CA2 (described in Section 2) is more suited to a generation of the synthetic pattern. Despite a high concentration, the starting pattern has the same islands number as the T2. On the other hand, the value of interface, $I_{initial}(Synth2) = 3428 > 3252$, is a clearly higher one than for T2. Moreover, the both positions of characteristic maxima, that is $k_{max}(S_\Delta; Synth2) = 26$ and $k_{max}(C_S; Synth2) = 30$ are different from those for the target, $k_{max}(S_\Delta; T2) = 32$ and $k_{max}(C_S; T2) = 35$. Thus, the starting situation is now much less comfortable compared to Example 1. Again, first we present, in Figs. 5b-d, typical reconstructions obtained by our method with the bottom, middle and top value of the $\alpha$-parameter for a chosen random seed. Once more, one can notice that the higher the $\alpha$ is, the smoother the shapes of the islands appear in the reconstructed patterns.

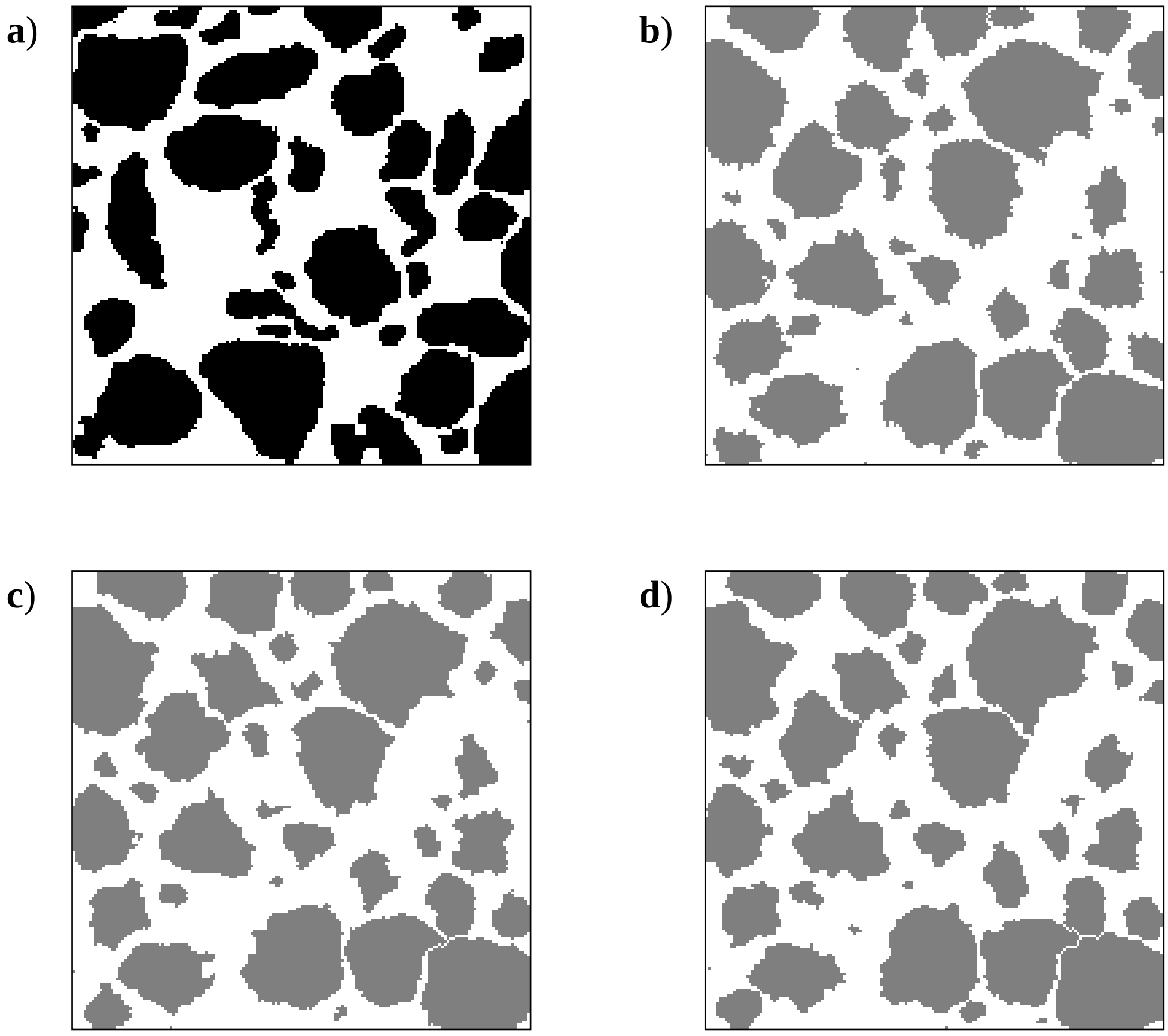

**Figure 5.** A comparison of the representative 170 × 170 digitized sub-domain of the concrete cross-section with the irregularly shaped stone phase adapted from Ref. [13] with its typical WDH reconstructions for a given random seed and the chosen set of $\alpha$-parameters. a) Target image T2; b) The reconstruction with $\alpha = 0.1$; c) $\alpha = 0.5$; d) $\alpha = 0.9$.

To show the quality of one of the 20 reconstructions using the example of the middle value of $\alpha = 0.5$ we present Figs. 6a-b. The related reconstruction R2*($\alpha = 0.5$) is shown in the inset of Fig. 6b. It corresponds to the closest $N_a(\alpha = 0.5)$ value to the average $<N_a(\alpha = 0.5)>$. As in the previous case, in Fig. 6a, the thick solid black lines (open and filled grey circles) correspond to entropic descriptors of spatial inhomogeneity and statistical complexity computed for the target T2 (reconstructed R2*) pattern, respectively. For completeness, the two normalized EDs obtained for the initial synthetic pattern Synth2, are described by the dashed lines (red and blue online). Respectively, in Fig. 6b, all the results for the CFs are shown. Although not so clearly as in Example 1, the curves for the Synth2 are quite close to the target ones. This is ensured by the inclusion of target's characteristic attributes to the creation procedure employed by the chosen cellular automaton. At all scales $k$, a significant fit of the considered hybrid descriptors can be observed. In the same way as in the previous example the small tolerance $\delta$-value, forces fitting of equal quality for other random seeds.

**6a**)

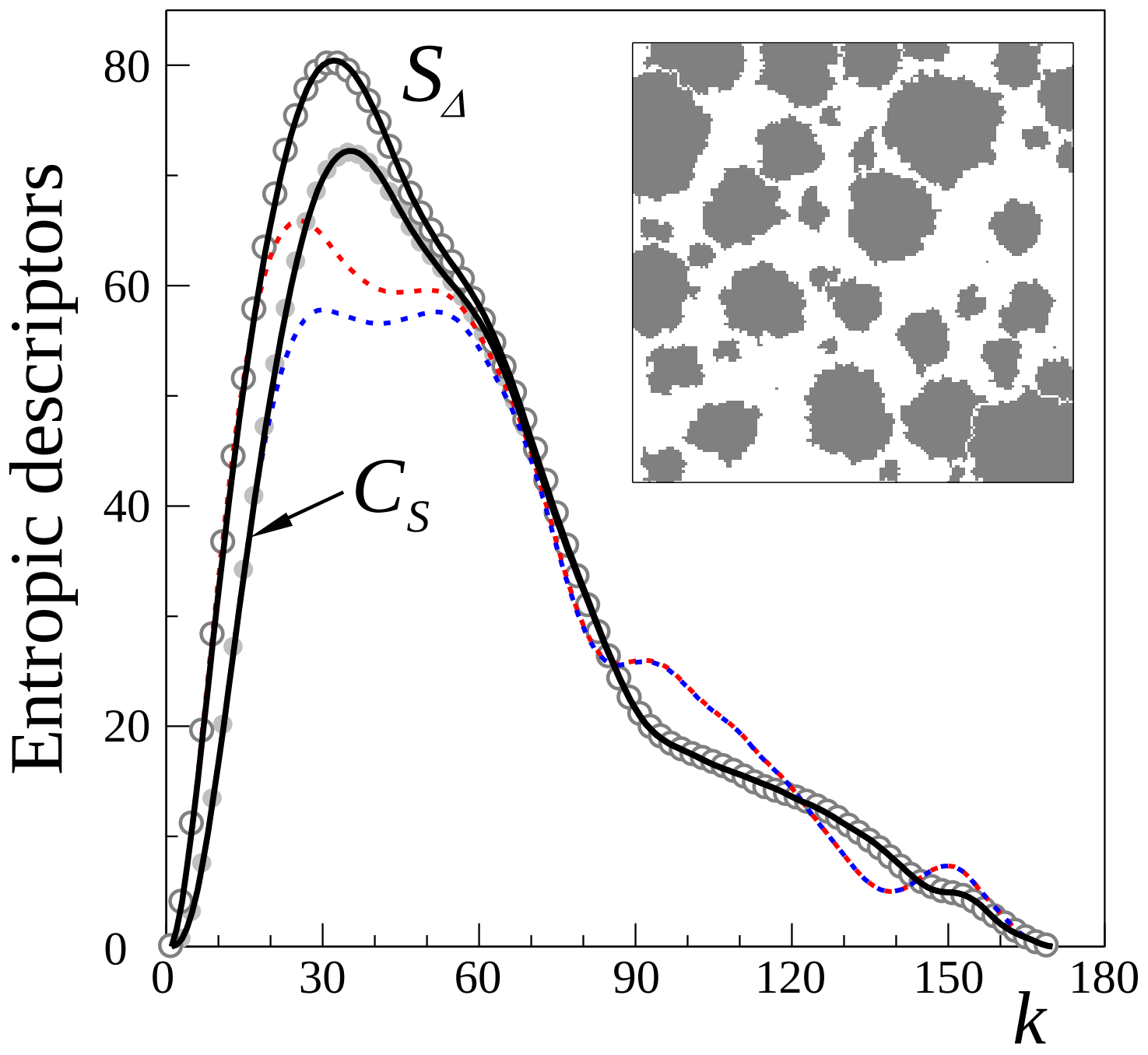

**6b**)

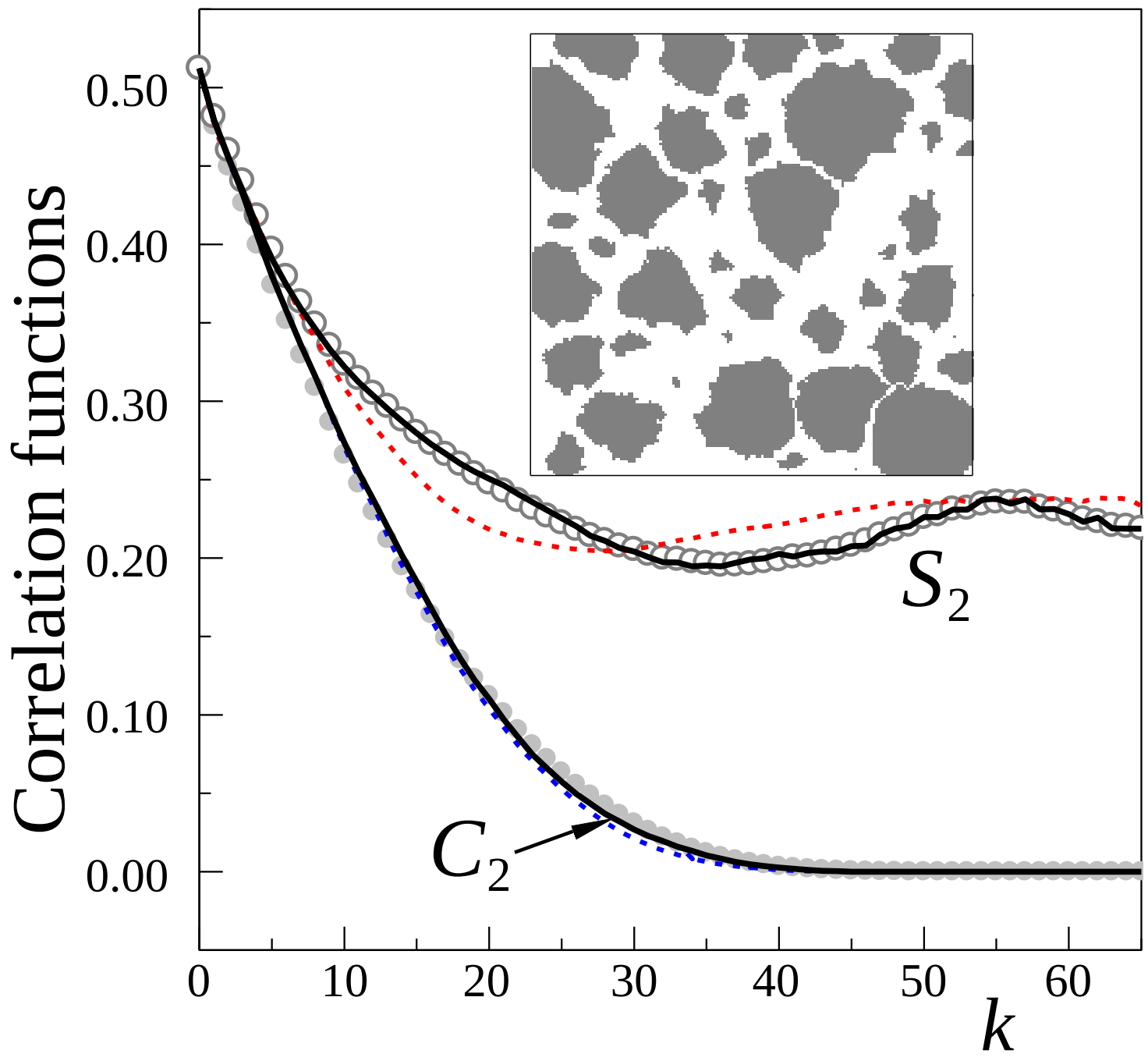

**Figure 6.** (Colour online) The quality of the weighted doubly-hybrid (WDH) approach using the example of the reconstruction R2*($\alpha = 0.5$; T2) selected as having the closest $N_a(\alpha = 0.5)$ value of the accepted MC steps to their average $\langle N_a(\alpha = 0.5)\rangle$ over the series of 20 runs. a) For the T2-target, thick solid black lines, and for its R2*-reconstruction, the open and filled grey circles relate to the pair of entropic descriptors $\{S_\Delta; C_S\}$. For completeness, the dashed lines, red and blue online, describe the computed starting values for initial Synth2-pattern that is shown in the inset; b) The same correspondence for the pair of correlation functions $\{S_2; C_2\}$ but now the R2*-reconstruction is presented in the inset.

Again, the statistics of the results for the completed series of 20 runs, each for the nine $\alpha$-parameters, is presented in Fig. 7. The dashed line connecting the averages $<N_a(\alpha)>$ of the accepted MC steps is a guide for eyes only. Once more, the shape of the curve is not a universal one. However, except for $\alpha = 0.5$ it shows a general decreasing trend in the $<N_a(\alpha)>$ belonging roughly to the range from 4000 to 6000. We point out that a standard MC reconstruction starting with a random initial configuration and using only the first terminating condition F1, after about $N_a = 2\times10^5$ steps did not achieve the needed accuracy of reconstruction. Thus, our approach indeed reduces the $N_a$ in the case of islands which are irregular in shapes.

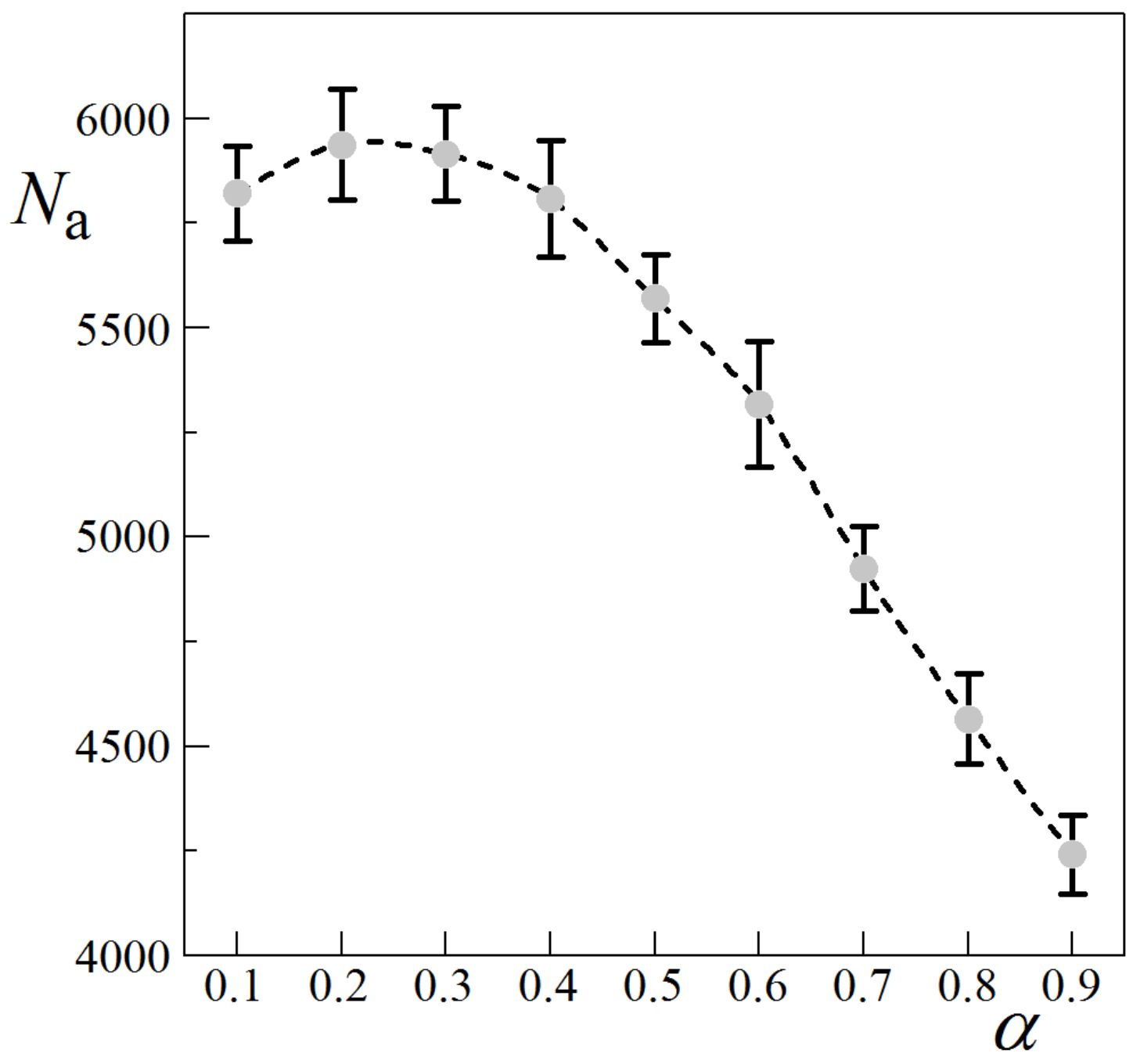

**Figure 7.** The statistics of $N_a(\alpha)$ accepted MC steps for the WDH reconstructions of the T2-target (given in Fig. 5a). The Synth2-pattern depicted in the inset of Fig. 6a is the initial configuration. For each of the fixed nine values of $\alpha$-parameters a series of 20 runs with given different random seeds was completed. The error bars represent the corresponding standard deviations. The dashed line connecting the averages $<N_a(\alpha)>$ marked by the filled circles serves as a guide for eyes only. Like previously, the shape of the curve is not a universal one. Now, for $\alpha > 0.2$ it shows the decreasing trend in the numbers $N_a$ which belong roughly to the range from 4000 to 6000.

The fast convergence of the WDH reconstruction for the three $\alpha$-cases given below and the closest to the averages $<N_a(\alpha)>$ is clearly demonstrated in Fig. 8. The curves from the top to the bottom (the thick green line for $\alpha = 0.9$, the medium blue one for $\alpha = 0.5$ and the thin red one for $\alpha = 0.1$ in colour on-line) refer to absolute values of the energy described by (3.3).

The arrows indicate the respective final values of $N_a(\alpha)$. The initial differences appearing in energies $E$ become negligible for each of $\alpha$-parameter when $N_a(\alpha) > 2000$. To illustrate the shrinking of the interface value to the target one, the dependence of $I_{current}(\alpha)$ on the number $N_a(\alpha)$ is depicted in the inset of Fig. 8. Similarly, as for the previous target T1, the dependence on $\alpha$-parameter rapidly becomes trivial, since for $N_a > 200$ the values of $I_{current}$ belong to the narrow range around the $I_{target}(T2) = 3252$.

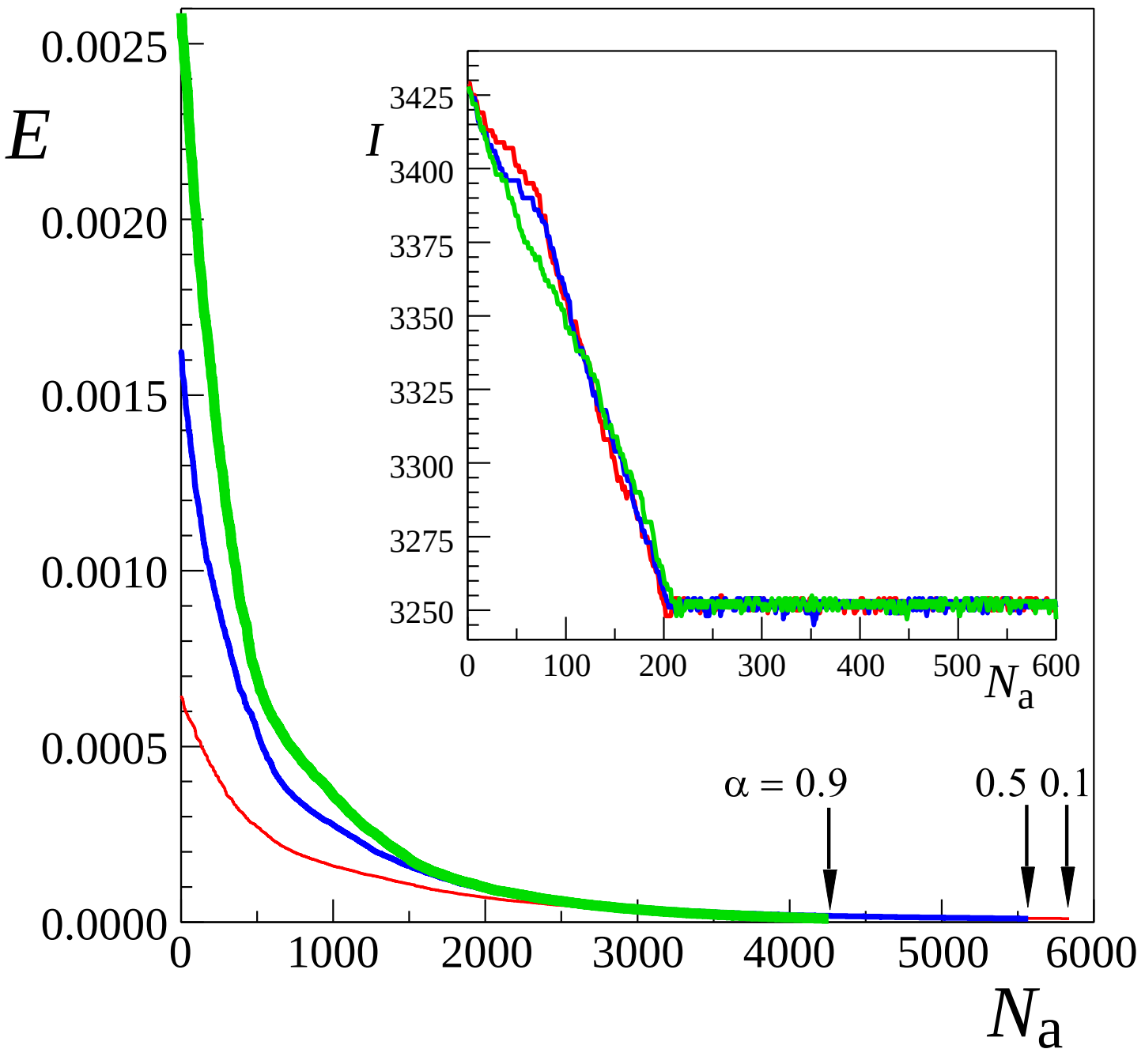

**Figure 8.** (Colour online) The convergence of energy $E$ as a function of the accepted Monte Carlo attempts $N_a$ for three representative cases: $\alpha$ = 0.9 (thick green), 0.5 (thin blue) and 0.1 (the thinner red line) from the top to the bottom, which are the closest to the averages $<N_a(\alpha)>$. The arrows indicate the corresponding final values of $N_a(\alpha)$, which are obtained when the assumed tolerance $\delta = 10^{-5}$ is achieved and the remaining two conditions (see the text) are fulfilled. To illustrate the limitative mechanism for the current interface value, the dependence of the $I_{current}(\alpha)$ on the $N_a(\alpha)$ is shown in the inset. In fact, for $N_a > 200$ one can see only the interface fluctuations around the $I_{target} = 3252$.

We expect our approach will be confirmed by further tests for patterns exhibiting complicated cluster structures. At this stage, the presented results support the reasonable strategy of speeding up of the microstructure reconstruction, at least for islands patterns. The last remark relates to the future application of the WDH method for random multi-phase materials, focusing on spatial dispersion for each of the phase components. This kind of possibility provides the method of the so-called phase descriptors introduced recently [27]. Those statistical descriptors were obtained as a result of splitting of the adapted overall entropic descriptor of a pillar model applied to greyscale images [28]. Each of the phase descriptors describes separately the corresponding contribution to the overall spatial inhomogeneity of the system. This allows studying how the spatial arrangement of each of the phases influences directly the outcomes of a MC reconstruction.

The meaning of the usage of distinct descriptors in the statistical reconstructing of microstructures has been recently reviewed in Ref. [29]. This review summarizes the descriptors and formulations used to represent structures at the microscale and mesoscale. In addition, the valuable comparison of the efficiency of reconstruction algorithms was given.

## 5. Summary

This work demonstrates the first implementation of our approach to MC reconstruction of microstructure of random heterogeneous media. The focus is given to patterns of islands of miscellaneous shapes and poly-dispersed in sizes, which are represented by two real samples. The weighted doubly-hybrid (WDH) method, besides the pair of two entropic descriptors, the $S_\Delta$ (for the spatial inhomogeneity) and the $C_S$ (for the statistical complexity), employs additionally a pair of two-point correlation functions, the $S_2$ (a standard correlation function) and the $C_2$ (a cluster one). The objective function given by (3.3) is a linear combination of two contributions: the first provided by the entropic pair and the second from the correlation pair. This allows some kind of adjusting of the weighting $\alpha$-parameter to a given type of microstructure of a target. In such a way, one can take into account partially a level of irregularity of the shapes of islands. On the other hand, the presence of the two hybrid pairs of different origin, the EDs and CFs, extends the range of observable additional spatial features. For instance, it may take into account those which are detected separately only by a member of one of the pairs. Thus, to minimize the summed squared differences in the objective function (3.3), a certain comprise is necessary for any of the fixed values of the tailoring $\alpha$-parameter. Hence, the two different and competing contributions do not facilitate obtaining of an assumed accuracy of reconstruction.

To speed-up the reconstruction process, instead of a standard random initial configuration, the synthetic one has been suitably prepared by a cellular automaton. The synthetic pattern shows some attributes characteristic of a given target medium. The performed simulations have revealed the $\alpha$-dependent trend in the average $\langle N_a(\alpha) \rangle$ of the accepted MC steps required for terminating the reconstruction for the given conditions. Further, the specific switching procedure between modified weak/strong bias modes has been implemented. This enhancement allowed tracing of the current values of the interface. In modelling of macroscopic properties of a heterogeneous medium such a tool is of some importance.

The utility of the WDH method was demonstrated by convincing and efficient reconstructing of microstructures for two adapted real samples. The approach provided credible enough shapes and sizes of islands, keeping their number and the target's interface. For the WDH reconstructions we have observed more regular shapes for higher values of $\alpha$-parameter; see Example 1. It suggests a kind of dominance of contribution of the entropic term in the cost function. Reversely, more irregular shapes need a lower value of $\alpha$. This time the contribution of the correlation term prevails as it has been shown in Example 2.

In conclusion, the computational cost in our approach is appreciably lower compared to the standard method. Moreover, any WDH-reconstructed pattern has the same interface and number of islands as the target. To the best of our knowledge, this is the first attempt at obtaining such a result. Further applications of the approach to complex multiphase materials are intended.

## Appendix

The general form of the entropic descriptors $S_\Delta$ and $C_S$ is

$$S_\Delta(k) = \frac{1}{\lambda}\,[S_{\max}(k) - S(k)] \tag{A.1}$$

and

$$C_S(k) = \frac{1}{\lambda}\,\frac{[S_{\max}(k) - S(k)][S(k) - S_{\min}(k)]}{[S_{\max}(k) - S_{\min}(k)]}\,. \tag{A.2}$$

The EDs make the use of micro-canonical current entropy $S(k) = k_B \ln \Omega(k)$, its maximum $S_{\max}(k) = k_B \ln \Omega_{\max}(k)$ and minimum $S_{\min}(k) = k_B \ln \Omega_{\min}(k)$, where Boltzmann's constant equals unity, $\Omega(k)$, $\Omega_{\max}(k)$ and $\Omega_{\min}(k)$ correspond to the number of microstates realizing the current, the most uniform and the most non-uniform configurational macrostate properly defined at length scale $k$. This scale is given by the side length of the sampling cell of size $k \times k$ sliding by a discrete unit distance. Black pixels are treated as finite size $1 \times 1$ objects. For instance, a current macrostate is described by a set of $i$th cell occupation numbers $\{n_i(k)\}$ by black pixels, $i = 1, 2, \ldots, \lambda(k)$. The number of allowed positions for the sliding cell equals $\lambda(k) = [L - k + 1]^2$. The detailed description of the spatial inhomogeneity entropic descriptor $S_\Delta$ can be found in [30] and the spatial statistical complexity $C_S$ in [31].